\documentstyle[12pt]{article}
\begin{document}
\newcommand{\beq}{\begin{equation}}
\newcommand{\eeq}{\end{equation}}
\newcommand{\beqn}{\begin{eqnarray}}
\newcommand{\eeqn}{\end{eqnarray}}


\gdef\journal#1, #2, #3, 1#4#5#6{		
    {\sl #1~}{\bf #2}, #3 (1#4#5#6)}		

\def\refstylenp{		
  \gdef\refto##1{ [##1]}			
  \gdef\refis##1{\item{##1)\ }}			
  \gdef\journal##1, ##2, ##3, ##4 {		
     {\sl ##1~}{\bf ##2~}(##3) ##4 }}

\def\refstyleprnp{		
  \gdef\refto##1{ [##1]}			
  \gdef\refis##1{\item{##1)\ }}			
  \gdef\journal##1, ##2, ##3, 1##4##5##6{	
    {\sl ##1~}{\bf ##2~}(1##4##5##6) ##3}}

\def\pr{\journal Phys. Rev., }

\def\pra{\journal Phys. Rev. A, }

\def\prb{\journal Phys. Rev. B, }

\def\prc{\journal Phys. Rev. C, }

\def\prd{\journal Phys. Rev. D, }

\def\prl{\journal Phys. Rev. Lett., }

\def\jmp{\journal J. Math. Phys., }

\def\rmp{\journal Rev. Mod. Phys., }

\def\cmp{\journal Comm. Math. Phys., }

\def\np{\journal Nucl. Phys., }

\def\pl{\journal Phys. Lett., }

\def\apj{\journal Astrophys. Jour., }

\def\apjl{\journal Astrophys. Jour. Lett., }

\def\annp{\journal Ann. Phys. (N.Y.), }

\def\jltp{\journal J. Low. Temp. Phys., }

\def\pmag{\journal Phil. Mag., }

\def\phtr{\journal Phase Trans., }

\def\jpA{\journal J. Phys. A, }

\def\jpC{\journal J. Phys. C, }

\def\ijmpA{\journal Int. Jour. of Mod. Phys. A, }

\title{Wess-Zumino-Witten model off criticality}
\vspace{3.5in}
\author{D. C. Cabra \thanks{CONICET-Argentina} \\
{\normalsize\it Departamento de F\'{\i}sica}\\
{\normalsize\it Universidad Nacional de La Plata}\\
{\normalsize\it C.C.67 - 1900 La Plata, Argentina}\\}
\date{}

\maketitle

\begin{abstract}
We study the renormalization group flow properties of
the Wess-Zumino-Witten model in the region of couplings between $g^2=0$
and $g^2=4\pi/k$,
by evaluating the two-loop Zamolodchikov's
$c$-function. We also discuss the region of negative couplings.

\end{abstract}
\newpage

\section{}
\ \indent  In this work we are going to study the Wess-Zumino model by
perturbation theory on a manifold with the topology
of the plane.

We construct
the Zamolodchikov's $c$-function from its original definition \cite{Zam}
up to two loops and show that it
fullfills the expected properties except the stationarity condition
at the trivial fixed point (FP).

In the region of
couplings between the trivial FP
$g^2=0$
and the well known infrared stable FP
$g^2=\frac{4\pi}{k}$ our $c$-function is monotonically decreasing
along a Renormalization Group trajectory (for large
values of $k$)  and takes
the value of the Virasoro central charge (VCC) of the critical
theory at the corresponding
FP's . It is stationary at the points $g^2=\frac{4\pi}{k}$ but it is
not stationary at the point $g^2=0$. In the region of negative
couplings between $g^2=-\frac{4\pi}{k}$ and $g^2=0$, where the theory is
non-unitary, the $c$-function is increasing to the infrared, takes the
values
of the VCC at the FP's,
is stationary at the point
$g^2=-\frac{4\pi}{k}$ but not at the trivial one.

Besides,
we show that our expression for the $c$-function coincides with
the one that can be obtained from the generalized definition
proposed by Cardy \cite{Cdy} for curved
manifolds
when applied to the Wess-Zumino model in $S^2$ \cite{LS}.
This is a non-trivial check since it is not proven in general that
the generalization of the $c$-function
to curved manifolds does fullfill the expected properties.

Using the result for the $\beta$-function obtained in
Ref.\cite{Bos} we show that the $c$-function is monotonically decreasing
to
the infrared and we find the particular form of the coefficient
which relates the $\beta$-function with the derivative of the
$c$-function
in our regularization scheme.
(This coefficient cannot be set equal to
one in an arbitrary regularization scheme \cite{BH}.) We find that in
our case this coefficient is positive definite.

The action for the Wess-Zumino-Witten (WZW) model is given
by \cite{Wi}:

\beq
W[h] = \frac{1}{2g^2}  \int d^2x tr\left(\partial_{\mu}h^{-1}
\partial_{\mu}h\right)+
 \frac{k}{12\pi}\int d^3y \epsilon_{ijk}tr\left( h^{-1}\partial_ih
h^{-1}
\partial_j h h^{-1}\partial_k h\right),
\label{1.1}
\eeq
where $h$ takes values in some compact Lie group $G$.

Besides the trivial fixed point, $g^2=0$,
we know for this model that it has an exact non-trivial fixed point
(IR stable) at :
\beq
g^2=\frac{4\pi}{k},
\label{1.2}
\eeq
as was stated in
Refs.\cite{Wi},\cite{KZ} using symmetry arguments.

Perturbative evaluations of the $\beta$-function have been done by
several
authors \cite{Bos}-\cite{chinos}, and the results
show the existence of this fixed point
at each order in perturbation theory.
Also, the VCC of the model at this point is known to be exactly:
\beq
c=\frac{kdim~G}{k+C_G}  ,
\label{1.2'}
\eeq
where $k$ is the level of the Kac-Moody algebra and $C_G$ is the
cuadratic
Casimir of $G$.

In the region of negative couplings, there is another point in which
the theory becomes conformally invariant \cite{Wi} which corresponds to:
\beq
g^2=-\frac{4\pi}{k},~~~~~k>0 ,
\label{1.2''}
\eeq
with VCC given by \cite{GK}:
\beq
c=\frac{-kdim~G}{-k+C_G}  .
\label{1.2'''}
\eeq
The corresponding theory is non-unitary and must be quantized with an
indefinite metric.

We will include this point in our study of the model since in the study
of coset models (both for bosonic \cite{GK}, \cite{K} and  fermionic
\cite{horda2} descriptions), WZW models with negative Kac-Moody central
charge do appear. (In those cases one has a BRST quantization condition
which avoids the appearance of negative norm states from the physical
spectrum \cite{K}.)

We are going to evaluate the Zamolodchikov's $c$-function perturbatively
using its original definition and
study the two regions corresponding
to $g^2$ in $[-4\pi/k,0]$ and $g^2$ in $[0,4\pi/k]$.

In order to make contact with Zamolodchikov's original construction, we
are going to evaluate:

\beqn
\lefteqn{c_{Zam}(g^2)=[2z^4<T(x)T(0)>} \nonumber \\
& &+4z^2x^2<T(x)\Theta(0)>-6x^4<\Theta(x)\Theta(0)>]
|_{x^2=R^2}, \label{1.12}
\eeqn
where $z=x_0+ix_1,~\overline{z}=x_0-ix_1$,
$T=T_{zz}$ and $\Theta=4T_{z\overline{z}}$ are the two independent
components of the energy momentum tensor in these coordinates
($R$ is a normalization point), by formulating the theory on a curved
world sheet in a generally covariant way.
We then have to take the functional derivatives
with respect to the background metric $\gamma_{\alpha \beta}$ and
finally
take the limit $\gamma_{\alpha \beta}\rightarrow \delta_{\alpha \beta} $.

To this end we define the effective action in a background metric as
usual:
\beq
e^{-S_{eff}[\gamma]}\equiv \int Dh e^{-W[h,\gamma]} ,
\label{1.13}
\eeq
and evaluate its finite part perturbatively up to two loops using
dimensional
regularization in an arbitrary metric \cite{Lu}.
In order to avoid infrared divergences one
must include a mass term \cite{Bos} but, as was shown in Ref.\cite{yo}
this
does not affect the $c$-function.

We just quote the result here (for details see Ref.\cite{yo}).

Our expression for the finite effective action is:
\beq
S_{eff}\{\gamma\}=(N^2-1)D\{\gamma\}\left[1-\frac{Ng^2}{8\pi}\left(3-
(\frac{g^2k}{4\pi})^2\right)+...\right],
\label{1.3}
\eeq
where:
\beq
D\{\gamma\}=\frac{1}{2}\left[ln~det(-\nabla^2)\right] .
\label{1.5}
\eeq
(The finite effective action has no essential
differences with the case
when the manifold has the topology of the sphere \cite{LS}.)

In the case at hand, the determinant of the laplacian operator is given
by \cite{Pol}:
\beqn
\lefteqn{D\{\gamma\}= -\frac{1}{48\pi}}\nonumber \\
& & \int d^2x d^2y \sqrt{\gamma(x)}
\sqrt{\gamma(y)}R(x)R(y)G(x,y)+ c\int d^2x \sqrt{\gamma(x)}
,
\label{1.14}
\eeqn
where $G(x,y)$ is the Green function of the covariant Laplacian:
\beq
\partial_{\mu}\left(\frac{1}{\sqrt\gamma}\gamma^{\mu\nu}\partial_{\nu}
\right)G(x,y)=\delta(x,y).
\label{G(x,y)}
\eeq

The connected part of the general correlator
of two energy momentum operators is defined through:

\beqn
\lefteqn{<T_{\mu\nu}(x)T_{\rho\sigma}(y)>-<T_{\mu\nu}(x)><T_{\rho\sigma}
(y)>=}
\nonumber \\
& &
-\frac{2}{\sqrt{\gamma(x)}}
\frac{2}{\sqrt{\gamma(y)}}\frac{\delta^{(2)}
S_{eff}[\gamma]}{\delta \gamma^{\mu\nu}(x)\delta \gamma^{\rho\sigma}(y)}
|_{\gamma^{\mu\nu}=\delta^{\mu\nu}}
\nonumber \\
& & +\frac{2}{\sqrt{\gamma(x)}}\frac{2}{\sqrt{\gamma(y)}}<
\frac{\delta^{(2)}
W[\gamma]}{\delta \gamma^{\mu\nu}(x)\delta
\gamma^{\rho\sigma}(y)}>|_{\gamma^{\mu\nu}=\delta^{\mu\nu}}.
\label{1.15}
\eeqn
where:
\beq
<T_{\mu\nu}>=\frac{2}{\sqrt{\gamma}}\frac{\delta}{\delta
\gamma^{\mu\nu}}
S_{eff}[\gamma].
\label{1.16}
\eeq

{}From eqs.(\ref{1.12})-(\ref{1.16}) it follows that (up to contact terms):

\beqn
<T(x)T(0)>&=&\frac{1/2}{z^4}c(g^2,k), \nonumber \\
<T(x)\Theta (0)>&=&0, \nonumber \\
<\Theta (x)\Theta (0)>&=&0 ,
\eeqn
where:
\beq
c(g^2,k)=(N^2-1)\left[1-\frac{Ng^2}{8\pi}
\left(3-\left(\frac{g^2k}{4\pi}\right)^2\right)
+...\right]
\label{1.7}
\eeq
and hence:

\beq
c_{Zam}\equiv c(g^2,k).
\label{1.17}
\eeq

It must be pointed out that there are no additional contributions
to these
quantities coming from the divergent part of the effective action
\cite{yo}.

Expression (\ref{1.17}) has the expected properties at the fixed points
given in eqs.(\ref{1.2}), (\ref{1.2''}). That is:
\beqn
c(g^2, k)\vert_{g^2=\pm\frac{4\pi}{k}}&=&(N^2-1)\left[1\mp\frac{N}{k}
+...\right] ,
\nonumber \\
\frac{\partial c(g^2, k)}{\partial g^2}\vert_{g^2=\pm\frac{4\pi}{k}}
&=&0 .
\label{1.18}
\eeqn

Since we are making a perturbative expansion with $1/k$ as a small
parameter,
we can study with the desired approximation the region of couplings
between
$g^2=0$ and $g^2=\frac{4\pi}{k}$. (We can also study separatedly
the region of negative
couplings between $g^2=-4\pi/k$ and $g^2=0$.)

We see that in the trivial fixed point, $g^2=0$, we have:

\beq
c(g^2, k)\vert_{g^2=0}=N^2-1 ,
\label{1.19}
\eeq
which corresponds to the value of the Virasoro central charge of $N^2-1$
free massless bosons, but the stationarity condition is not fulfilled at
this point:
\beq
\frac{\partial c(g^2, k)}{\partial g^2}\vert_{g^2=0}\neq 0 .
\label{1.20}
\eeq

In order to study the renormalization group behaviour of the
$c$-function we use
the result for the $\beta$-function quoted in Ref.\cite{Bos}:
\beq
\beta(g^2)=-\frac{Ng^4}{2\pi}\left[1-\left(\frac{g^2k}{2\pi}\right)^2
\right]+...
,
\label{1.21}
\eeq
which shows that $g^2(\mu)$ is increasing to the IR
so the Zamolodchikov's $c$-function
is decreasing along a RG-trajectory for $g^2 >0$ as expected.

If $g^2$ is allowed to take negative values then our expression for the
$c$-function shows that it is increasing to the IR, and it is stationary
at the non-trivial fixed point $g^2=-4\pi/k$. (The increasing of the
$c$-function does not contradict the $c$-theorem since for negative
couplings
the model is
non-unitary and hence Zamolodchikov's proff does not hold.)

We can also calculate the coefficient
which relates the $\beta$-function with the derivative of the
$c$-function,
$\beta(g^2)=F(g^2)\frac{\partial c(g^2, k)}{\partial g^2}$,
which up to this order is given by:

\beq
F[g^2]=\frac{2}{3}g^4(N^2-1)^{-1} ,
\label{1.22}
\eeq
and is positive definite as predicted in Ref.\cite{BH}. This explains why
for $g^2=0$ the vanishing of the $\beta$-function does not necessarily
lead to the
stationarity of $c$.

As mentioned in the introduction, in Ref.\cite{Cdy}, a generalization of
the
$c$-theorem to curved manifolds
was proposed. In this paper it has been suggested that a natural
definition,
which could be useful in the generalization of the c-theorem to four
dimensions, is given by:
\beq
\tilde{c}=-3\int_{S^2} <\Theta>\sqrt{\gamma}d^2x
\label{1.10}
\eeq
where the integration is done over the sphere $S^2$ endowed with the
metric
induced by embedding into $R^3$.

The v.e.v $<\Theta(x)>$
has been evaluated exactly in \cite{LS} at the fixed point
(\ref{1.2}).
In order to evaluate it
off criticality one has to look not only at the finite
part of the effective action, but also at its divergent part (which
behaves
as
$1/d-2$, where $d$ is the dimensionality of space-time), since it can
give a non-trivial, finite contribution to the trace.
However, in this case it is easy to show that  the trace has no
additional
contributions arising from the divergent terms in the effective
action.
The result is simply:
\beq
<\Theta>\equiv \gamma^{\mu \nu}(x)<T_{\mu \nu}(x)>=c(g^2,k)
\frac{R(x)}{24\pi},
\label{1.6}
\eeq
where $R(x)$ is the scalar curvature, and $c(g^2,k)$ is given by eq.(15).

We then have simply that:
\beq
\tilde{c}\equiv c(g^2,k).
\label{1.11}
\eeq

This result must be taken with a grain of salt since there does
not exist a complete proof that this generalized "c-function" fulfills
the conditions that the Zamolodchikov's c-function does. In particular
the decreasing property of this function has not been proven, although
it was
verified to lowest order in perturbation theory \cite{Cdy}.
However, we see that the explicit calculation of the Zamolodchikov's
$c$-function, (eq.(16)), coincides with the one obtained from
(\ref{1.10}).

As a final comment it is interesting to note that the finite effective
action
can be written as:
\beq
e^{-S_{eff}[\gamma]}=\left[det(-\nabla)\right]^{-\frac{1}{2}c(g^2,k)},
\label{1.23}
\eeq
which is the $c(g^2,k)$-power of the partition function for a massless
free
boson.
This fact fits with the intuition that the $c$-function is a measure of
the massless degrees of freedom.

I would like to thank C.Na\'on, E.Moreno and G.Rossini for helpful
discussions and for carefully reading the manuscript.


\begin{thebibliography}{99}
\bibitem{Zam} A.B.Zamolodchikov, {\sl Sov.Phys.JETP Lett.}{\bf 43}, 731,
(1986).
\bibitem{Cdy} J.L.Cardy, \pl 215B, 749, 1988.
\bibitem{LS} H.Leutwyler and M.Shifman, {\sl Preprint BUTP-91/6,
March 91}.
\bibitem{Bos} M.Bos, \pl 189B, 435, 1987.
\bibitem{BH} D.Boyanovsky and R.Holman, \prd 40, 1964, 1989.
\bibitem{Wi} E.Witten, \cmp 92, 455, 1984.
\bibitem{KZ} V.G.Knizhnik and A.B.Zamolodchikov, \np B247, 83, 1984.

\bibitem{chinos} Zheng-Min Xi, {\sl CCAST preprint}, 1988.

\bibitem{GK} K.Gawedzki and A.Kupiainen, \pl 215B, 119, 1988.
\bibitem{K} D.Karabali, Q-Han Park, H.Schnitzer and Z.Yang, \pl 216B,
307,
1989.
\bibitem{horda2} K.Bardacki, E.Rabinovici and B.Saring, \np B229, 151,
1988;
D.Cabra, E.Moreno and C.von Reichenbach, {\sl Int. J. Mod. Phys. A}
{\bf 5},
2313, (1990).
\bibitem{Lu} M.Luscher, {\sl Ann.Phys. (N.Y.)}{\bf 142}, 359, (1982).

\bibitem{yo} D.C.Cabra, {\sl Phys.Rev.D, in press}.
\bibitem{Pol} A.M.Polyakov, \pl 103B, 207, 1981.
\end{thebibliography}
\end{document}